# Experimental Study of Energy Transfer by Inertial Waves During the Build up of Turbulence in a Rotating System

Itamar Kolvin, Kobi Cohen, Yuval Vardi and Eran Sharon
The Racah Institute of Physics, The Hebrew University of Jerusalem
erans@vms.huji.ac.il

## Abstract

We study the transition from fluid at rest to turbulence in a rotating water cylinder. We show that the energy, injected at a given height, is transported by inertial wave packets through the fluid volume. These waves propagate at velocities consistent with those calculated from linearized theory[1], even when they possess large amplitudes. A clear "front" in the temporal evolution of the energy power spectrum is detected, defining a time scale for energy transport at the linear wave speed in the system. Nonlinear energy transfer between modes is governed by a different time scale that can be much longer than the linear one. These observations suggest that the energy distribution and statistics in rotating turbulent fields that are driven by intermittent energy sources may be different from those described by the inverse energy cascade in two-dimensional turbulence.

The physics of flows in rotating systems are important for the understanding of astrophysical, atmospheric, geophysical and engineering systems. It is well established that when the rotation of the system is high enough, a flow becomes quasi two-dimensional (quasi 2D)[2,3]. In particular, rotation has a dramatic effect on turbulent flows in three-dimensions (3D), converting them into quasi 2D turbulent fields. Such steady[4], as well as decaying[5], flows have been extensively studied, with the similarities and dissimilarities of their characteristics to those of 2D turbulence noted in both experimental [6] and theoretical [7] studies. Much less work has been dedicated to studying the evolution of rotating turbulence, its response to abrupt changes in energy injection rates, and to the processes governing energy and momentum transfer in the medium. These questions are of importance when considering the dynamics and statistics of natural flows, such as atmospheric flows, that are, typically, driven by a fluctuating energy source.

Consider a flow field in a 3D volume that rotates with an angular velocity $\Omega$ ($\Omega=\Omega_z$). For typical velocity and length scales U and L, and viscosity $\eta$, there are two dimensionless numbers that characterize the flow. The Reynolds number, $\mathrm{Re} \sim \frac{UL}{\eta}$, which quantifies the ratio between the magnitudes of the inertial and viscous terms in the Navier-Stokes (N.S.) equation, and the Rossby number $Ro \sim \frac{U}{2L\Omega}$, which quantifies the ratio between the inertial and Coriolis terms. When Re>>1 and Ro<<1 Coriolis acceleration dominates both inertial and viscosity terms. Under these conditions, gradients of the velocity field parallel to $\mathbf{\Omega}$ are small and in this sense the flow becomes quasi 2D. In the limit Ro→0, the inertial term is completely neglected and the equation of motion becomes linear. In this approximation, the Coriolis force balances the pressure gradients and generates a finite restoring force that resists "deformations" of the velocity field in the z direction[1,8]. The velocity field, thus, becomes "elastic" and supports the propagation of inertial waves. Such waves were derived for the case of small and slow perturbations to a fluid that rotates as a solid body (fluid at rest in the rotating system) and the magnitude of their group velocity was shown to be given by $|v_g(k)|=2\Omega/k$, where $\boldsymbol{k}$ is the wave number[8, 1]. Inertial waves emitted from a weak point source were observed experimentally [9] and numerically [10] and the traces of their resonant modes were recently measured in decaying turbulent fields in a rotating tank [11].

It was recently suggested [12] that nonlinear resonant interactions of inertial waves, that transfers energy to large scales, is responsible for the build up of a turbulent field with a $k^{-3}$ energy spectrum at large scales. Such an energy spectrum is inherently different from the $k^{-5/3}$ spectrum, obtained from the inverse energy cascade (local in k) in 2D turbulence. However, inertial waves were derived within a linearized theory and studied only for small (and slow) localized perturbations to fluid at rest. The propagation of inertial wave packets within an existing, energetic flow field has never been measured. It is an open question of how and how much energy they can transfer to the entire fluid volume before nonlinearities destroy them. We suggest that these issues are of special importance when considering rotating systems that are driven by energy bursts.

In this work we present direct measurements of inertial waves in a rotating system, during the transition to turbulence. We show that depending on the system's parameters, these waves can propagate to large distances before nonlinear effects take place. During these times, the waves serve as the *main* mechanism of energy transport parallel to the axis of rotation, and their *linear* dispersion relation is maintained even at large Reynolds numbers (of order $10^4$) and Rossby numbers Ro~1. We further demonstrate that the evolution of the flow field towards a quasi 2D turbulent field is dominated by two different time scales, one that characterizes the spatial transport of energy by inertial waves, and a second that is associated with nonlinear energy transport to large spatial scales.

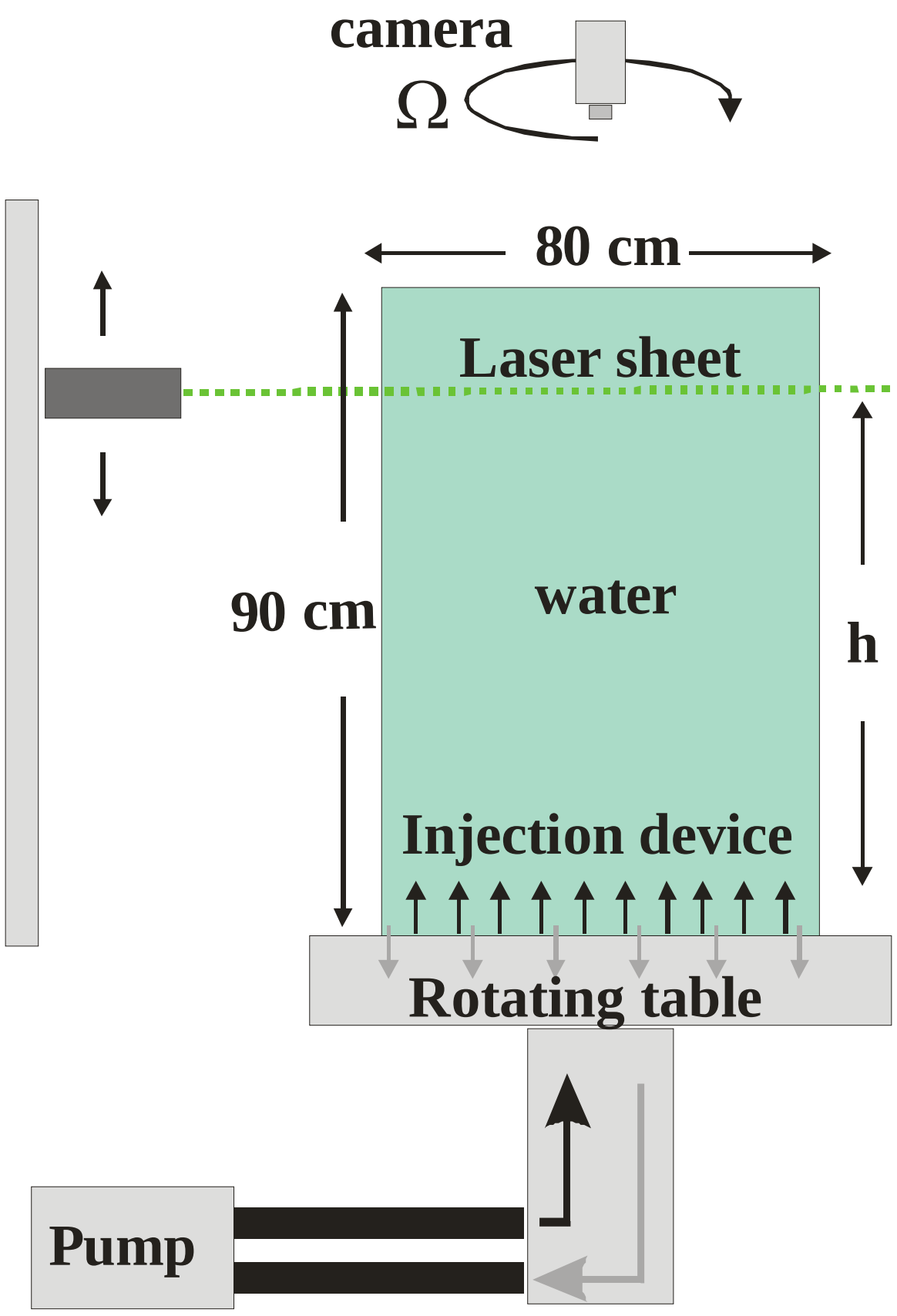

**Figure 1:** The experimental system. A covered, cylindrical transparent water tank is placed on a rotating table (Ω max= 16 Rad/sec.). Water is injected and drawn in through the injection device (see text) at the bottom of the tank. Laser sheets illuminate a cross section at a variable height, *h*, and a co-rotating camera grabs images of the light scattered from 50 μm diameter Polyamide particles suspended within the water.

Our experimental system (Fig. 1) consists of a Plexiglas cylinder, of 80 cm diameter and 90 cm height (the z direction), placed on a rotating table ( $\Omega=\Omega_z$ up to 16 Rad/s). The tank is filled with water and covered with a transparent flat lid that serves as the upper boundary of the fluid volume. The flow is driven by an injection device that covers the entire base of the cylinder. It consists of 248 flexible silicon tubes outlets (0.8 mm diameter) arranged every 4 cm in a hexagonal grid and 73, 6 mm diameter inlets arranged in a hexagonal grid, with 7 cm spacing, that overlaps the outlet grid. A 2.5 KW positive displacement pump circulates the water, at flow rates, *Q*, up to $Q_{max}=3\ l/s$, resulting in a maximum injected power of $P_{max}= 300\ W$. The distribution of the sources results in energy injection at a central wavelength of about 5 cm with negligible power at larger scales. The water is seeded with 50 μm diameter Polyamide particles, and sections of the cylinder at variable height, *h*, from the bottom (*10<h<80 cm*), are illuminated with two 1.5 W horizontal laser sheets of 3 mm thickness. Images of the light scattered from the particles, are acquired by a 30 fps, 1Mpix camera rotating with the system. These images are used for PIV measurements of the horizontal velocities within the illuminated section. Strobing the laser in synchronization with the camera allows shortening of the effective times between two PIV frames to below 1ms.

At high rotation rates, a region of quasi 2D flow, with negligible divergence in the *x,y* plane, is formed at the upper part of the cylinder (where Ro is the smallest) [14] . When this region is continuously fed by energy injected from the bottom of the tank, it evolves into a quasi 2D turbulent field, similar to previously studied rotating steady-state flows generated by a vibrating grid [4] and other energy injection devices [13, 14]. In the current work we focus on the early stages of the transition from solid body fluid rotation to turbulence. To study this transition, the system is first brought to a solid body rotation. The injection system is then activated (*t=0*) at a set flow rate. Starting at *t=0*, PIV measurements are performed at a given height, *h*. The measured horizontal velocity fields, $(u(x,y), v(x,y))$, are analyzed to obtain the vorticity, kinetic energy density and energy power spectrum at successive times *t*.

Figure 2 presents snapshots of the kinetic energy density, $E(x,y)=u^2(x,y)+v^2(x,y)$, measured at height *h=56 cm*, for increasing *t*. The fluid is static at small *t* (Fig. 2 a) and it is only after a delay of ~ 3 s that flow is initiated at *h* over the entire cross section (Fig. 2 b). As time progresses (Fig. 2 c), the intensity of the flow increases, while the typical spatial scales remain small (as in Fig 2 b). The flow field continues to vary slowly (Fig. 2 d) and only after time, *t*, of order $10^2$ s is a steady turbulent flow obtained (Fig. 2 e).

The time evolution (Fig. 3) of the (azimuthally averaged) energy power spectrum, $E(k)$, at *h* has a distinct signature. The energy is not populated simultaneously at all injected wave numbers. Instead, two different types of "fronts" (in the *k-t* plane) are responsible for the transport and distribution of the turbulent energy. The first type of front (dotted line) indicates that, although the entire spectrum was injected simultaneously, each wave number, *k*, has a distinct (and different) arrival time at a given plane in *z*. The arrival time, $\tau_1(k)$, increases linearly with *k*. This initial stage spans a few seconds, after which the variations in the spectrum, as well as in the total energy, are much slower. $\tau_1(k)$ is defined as the time at which the energy entrained in each mode, *E(k)*, increases at the highest rate (see inset of Fig. 3 for $k=0.17$ $cm^{-1}$).

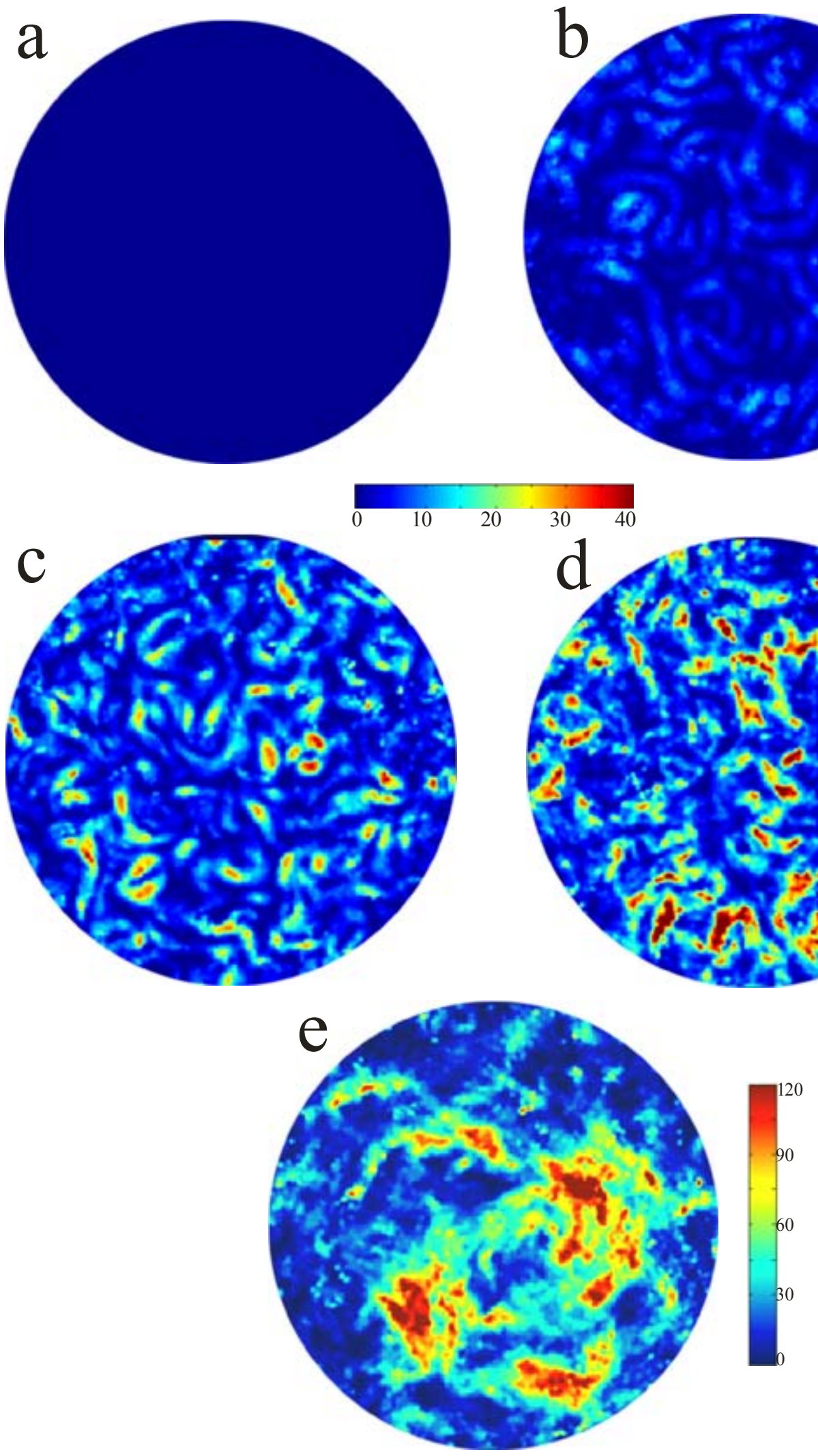

**Figure2:** Five snapshots of the kinetic energy density field, $E=v^2+u^2$, measured at different times (*t=1.3 , 3.3, 4.7, 10 and 100 s.* for **a-e**) after the initiation (*t=0)* of energy injection. Energy appears over the entire cross section at *t=3.3 s* (b). The amount of energy increases, while the typical length scale slightly decreases (c). Some rearrangement of eddies is visible at *t =10* s (d) and it is only after time of order tens of seconds that a steady turbulent field is achieved (e). The measurements were obtained for *h=56 cm*, Ω=9.4 Rad/s and P= 6 W. Each cross section is of 80 cm diameter. A linear color bar in $(cm/s)^2$ is used for a-d, whereas in (e) its range is expanded by a factor of three.

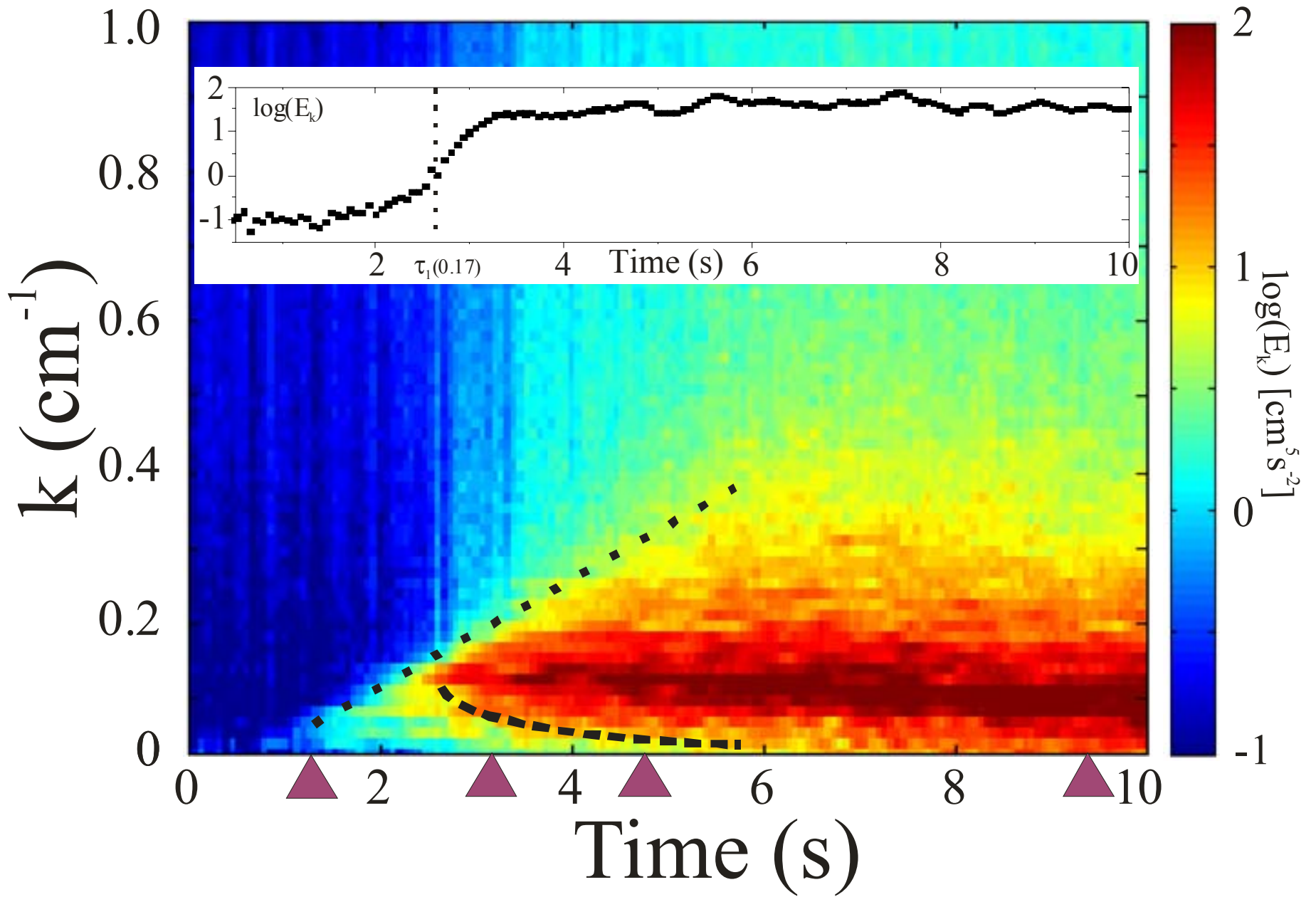

**Figure.3:** A typical plot of the temporal evolution of the energy power spectrum. A color map of the energy spectrum, *E(k)* (logarithmic color bar), as a function of the time, *t*, from the initiation of energy injection. The spectrum does not contain energy for the first few seconds, after which a linear front is observed (marked with a dotted line), indicating a distinct time, $\tau_1(k)$, for the "arrival" of each wave number *k*. The inset shows the time evolution of *E(k=0.17)*. The energy increases within one second, defining a time $t= \tau_1(k)$ (marked with dashed line) after which it varies much more slowly. A second front can be observed at small wave numbers (dashed line), defining a second time $\tau_2(k)$, which decreases with *k* and indicates nonlinear energy transfer from small to large scales. The purple triangles mark the times at which the flow fields presented in Fig 2 (a)-(d) were measured. Experimental conditions were as in Fig. 2.

We have conducted more than forty experiments over a range of heights ( $32\text{ cm} < h < 73\text{ cm}$), rotation rates ($6.3\text{ Rad/s} < \Omega < 14\text{ Rad/s}$) and energy injection rates ($1.5\text{ W} < P < 45\text{ W}$). Fronts similar to those presented in Fig. 3 appeared in all of these measurements. Plotting $\tau_1(k)$ versus *k* for various heights and rotation rates yields a set of quasi linear curves (Fig. 4 a), from which it is observed that $\tau_1(k)$ decreases with $\Omega$ and increases with *h*, for each *k*. Fig. 4b shows that scaling $\tau_1$(k) by $h/2\Omega$ leads to a data collapse onto a linear curve of slope 0.9 ± 0.05 (Fig. 4b). Since the group velocity of inertial waves is $v_g(k)=2\Omega/k$, the collapse implies that $\tau_1(k)$ is the *traveling time* of waves from the source to the measurement plane at a velocity dictated by the *linear* dispersion relation. Thus the observed front in Fig. 3, marks the arrival of planar inertial waves emitted from the injection plane, to a given measurement plane. In our experiments these planar wave-fronts travel in the range 30< $v_g(k)$<120 cm/s.

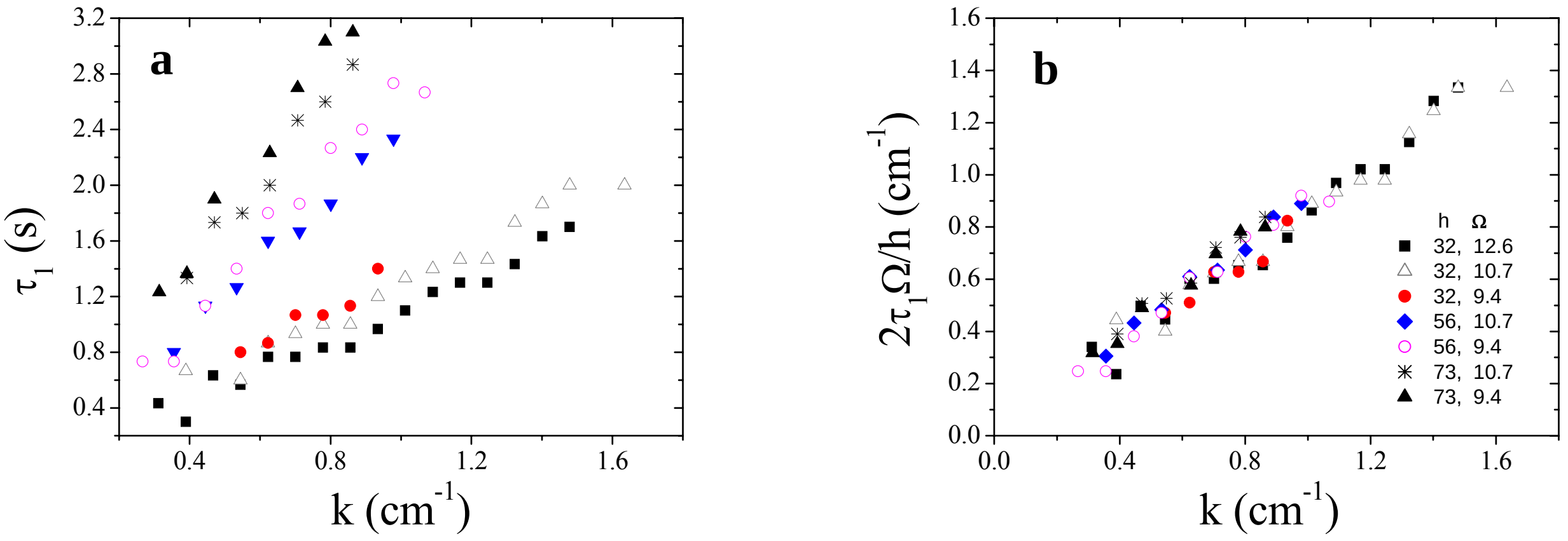

**Figure 4:** The arrival time of inertial waves. a) $\tau_1(k)$, measured for various heights and rotation rates yielding a set of quasi-linear curves. b) Once $\tau_1$ is rescaled by $h/2\Omega$, these data collapse onto a linear line of slope 0.9 ± 0.05. The front is, thus, consistent with the traveling time of waves with group velocity $v_g=2\Omega/k$ from the injector to the measurement plane. The measurement heights (in cm) and rotation rates (in Rad/s.) are indicated in (b). Energy injection rate: *P=12 W*.

In Figure 5 we fix *h* and $\Omega$ and vary the energy injection rate, *P*, over the range *1.5-45* W. Even though the injected power spans over a decade, $\tau_1$ is unaffected, an additional indication of the linear nature of the waves. This linear nature is somewhat surprising. The flow fields generated by the waves have high Reynolds numbers (estimated using the velocity rms and the central injected length scale) Re~ $10^3$, which are typical to turbulent flows. In addition, Ro~1, so it is far from obvious that nonlinear terms in the NS equation can be neglected. We also see that shorter waves propagate within a medium which is already excited by the longer waves. For example, at *t=3* s, where Re of the flow in Fig.3 is of order $10^3$, the velocity of the high *k* modes is unaffected by their propagation through a medium that is already highly excited by the lower modes. These waves can, therefore, *not* be regarded as small and slow perturbations to a static fluid, but as a more general mode of energy transport.

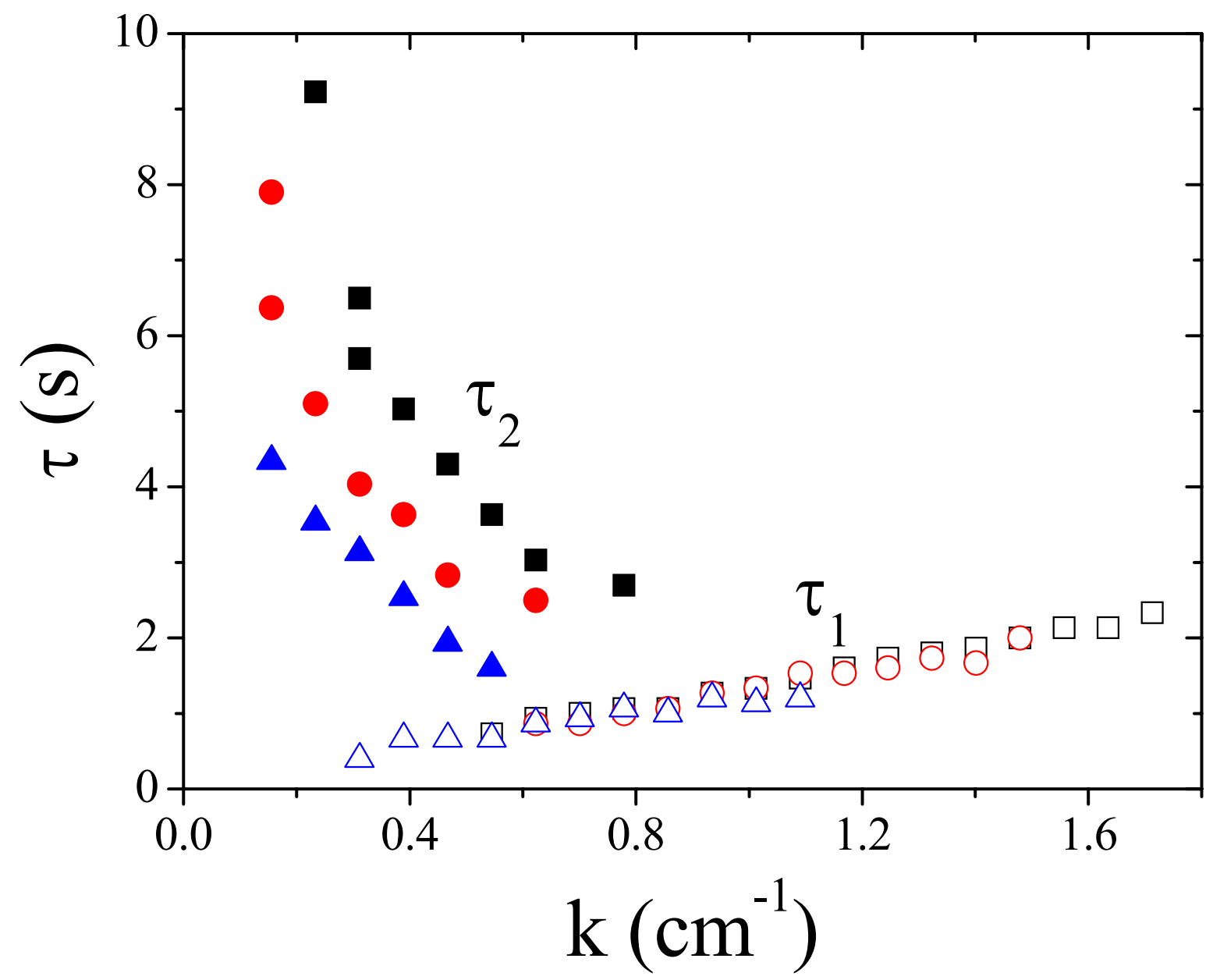

**Figure 5:** The effect of energy injection rate on $\tau_1$ and $\tau_2$. $\tau_1$ (open symbols) and $\tau_2$ (full symbols) versus the wave number *k*, measured at *h=32* cm, *Ω=11* Rad/s and different injection rates P (squares- 1.5 W, circles- 6 W, triangles- 45 W)). The higher the power injected, the shorter $\tau_2$ is, as expected from a time that is governed by nonlinear interactions. On the other hand, the linear time, $\tau_1$(k), shows no dependence on P.

Once the front arrives at a given plane, *sustained* turbulent motion develops. In order for turbulence to evolve beyond the conditions transported by the front arriving at time $\tau_1$, nonlinear mode interactions must occur. Nonlinear energy transfer is indicated by a second "front", indicated by the dashed line in Fig 3, at which *E*(*k*) increased sharply with time. We define a characteristic, mode-dependent time, $\tau_2(k)$, marking the increase in *E*(*k*). $\tau_2$ decreases with the wave number and is strongly dependent on the Re (Fig. 5), thus indicating the nonlinear nature of the underlying mechanism for wave interactions. This second “nonlinear” front marks the initial stages of the energy transfer from small to large scales. Indeed, the typical, "nonlinear time" for energy transfer across the spectrum – the eddy turnover time – should scale as $(\mathrm{v}k)^{-1}$. Our measurements indeed show (Fig. 5) that $\tau_2(\mathrm{k})$ decreases with both $k$ and v, where v is taken to be the velocity rms. More detailed measurements are required in order to find the exact functional dependence of $\tau_2(\mathrm{k})$ on $k$ and v in our system. Note that even at times much longer than $\tau_2$, $E(k,t)$ still evolves in time and the steady-state spectrum (e.g. Fig. 2e) is only attained at a much longer time scale that is associated with the development of equilibrium across the energy spectrum. This time scale has to be determined by further study.

We, therefore, see that the energy transfer during the initial stages of the build up of the turbulent field is characterized by two independent time scales, $\tau_1(k)$ and $\tau_2(k)$. $\tau_1(k) \sim h/\mathrm{v}_g(k)=hk/2\Omega$ is amplitude-independent and associated with the propagation of inertial waves at the *linear* group velocity from the energy source to the measurement plane. The nonlinear time, $\tau_2(\mathrm{k})$, is associated with energy transfer across the spectrum and is expected to scale as $\tau_2(k)\sim(\mathrm{v}k)^{-1}$. Thus, when $k<(\Omega/h\mathrm{v})^{1/2}$ we have $\tau_1(k)< \tau_2(k)$, and the flow will be dominated by linear inertial waves, with no significant nonlinear effects. The range of this region scales like $h\sim\Omega k^{-2}\mathrm{v}^{-1}$ and, in principle, could be extended by tuning the parameters of the system. Within this range, inertial waves are the sole energy transport mechanism and temporal variations of the energy source will be preserved, as the waves carry them through the medium.

In summary, linear inertial waves are both capable of carrying large amounts of energy over large distances, and persist for long times (in terms of the rotation period). The dynamics of these waves should be taken into account when considering turbulent build up, or when studying rapidly rotating systems, driven by an intermittent energy source. In such systems, temporal and spectral properties of the energy source will be transported to long distances by inertial waves and will only change for times longer than $\tau_2$. It is known [17] that a $k^{-5/3}$ spectrum can be observed in rotating turbulent flows. It, however, is still an open question of whether this spectrum *always* develops. The above observations suggest that when the typical time of fluctuations in energy injection is short compared with $\tau_2$ and $\tau_1$, the steady state energy spectrum may well be influenced by both the propagation of inertial waves and their subsequent nonlinear interactions.

**Acknowledgments:** This work was supported by The Israeli Science Foundation, grant # 488-04.